\documentstyle[preprint,pra,prl,aps,epsf,axodraw]{revtex}
\begin{document}
%\draft
\title{The Standard Model in Strong Fields:
Electroweak Radiative Corrections for Highly Charged Ions.}
\author{Ilya Bednyakov, Leonti Labzowsky}
\address{Institute of Physics, St.Petersburg State University,
198904 Uljanovskaya 1, Petrodvorets, St.Petersburg, Russia}
\author{G\"unter Plunien, Gerhard Soff}
\address{Institut f\"ur Theoretische Physik, Technische  Universit\"at
Dresden, Mommsenstrasse 13, D-01062  Dresden, Germany}
\author{Valentin Karasiev}
\address{Instituto Venezolano de Investigaciones Cientificas
Caracas, Venezuela}
\date{\today}
\maketitle

\begin{abstract}
Electroweak radiative corrections to the matrix elements 
$\langle ns_{1/2}|{\hat H}_{\rm PNC}|n'p_{1/2}\rangle$ 
are calculated for highly charged hydrogenlike ions.
These matrix elements constitute the basis for the description of the most 
parity nonconserving (PNC) processes in atomic physics. 
The operator ${\hat H}_{\rm PNC}$ represents the parity nonconserving 
relativistic effective atomic Hamiltonian at the tree level. The deviation of 
these calculations from the calculations valid for the momentum transfer 
$q^{2}=0$ demonstrates the effect of the strong field, characterized by the 
momentum transfer $q^{2}=m_{e}^{2}$ ($m_{e}$ is the electron mass). 
This allows for a test of the Standard Model in the presence of strong fields 
in experiments with highly charged ions.
\end{abstract}
 
\section{Introduction}
%\bigskip
PNC experiments in atomic physics provide an important possibility 
to deduce
informations on the Standard Model independent 
of high-energy physics experiments. The recent LEP
experiments \cite{l1}, that yield extremely accurate values for Z-boson 
properties, correspond to the resonant process. This means that
all the nonresonant corrections are strongly suppressed. Hence this
might not be the most convincing way for the search of all types of 
``new physics'' beyond the minimal Standard Model, e.g.,
for the existence of a second Z-boson etc.

The observation of ``new physics`` in atomic physics experiments is most 
probable related to processes beyond the tree level, for which electroweak 
radiative corrections are taken into account. This question was thoroughly 
discussed by many authors, in particular in \cite{l2} - \cite{l6}. 
Atomic PNC experiments are usually performed with heavy atoms (Cs, Tl, Pb, Bi) 
due to the strong enhancement of PNC effects with increasing nuclear charge 
number $Z$, which was first noticed by Bouchiat and Bouchiat \cite{l7}. 
However, the electrons involved in various PNC processes in these atoms are 
loosely bound valence electrons and the corresponding momentum transfer is 
much smaller than the squared rest mass of the electron:
\begin{equation} 
 q^{2}\ll m_{e}^{2}
\;.
\end{equation}
The situation is different in highly charged ions (HCI), where 
$q^{2}\approx m_{e}^{2}$. This peculiar property of HCI was a reason 
for many experimental and theoretical efforts to ``test'' QED in strong 
fields. In particular experimental \cite{l8} - \cite{l11} 
and theoretical \cite{l12} - \cite{l17} attempts were made 
recently to verify the 
Lamb shift in one-electron heavy ions up to the second order in the fine 
structure constant $\alpha$. These attempts are yet uncomplete from both 
sides.

Therefore there would be even a deeper reason to test the validity of 
the Standard Model in strong fields. We should emphasize here that the 
concept of the strong field is not constrained
to the large momentum transfer in some particular processes. 
The latter often occurs in high energy physics collision experiments. 
During the collision period the particles are also influenced by the strong 
field. However, for the tightly bound electron in HCI this strong field is 
present for a relatively long time, defined by the lifetime of the 
corresponding electronic state.

In this paper we show that the test of the Standard Model (or the search 
for ``new physics'' beyond the Standard Model) is possible in experiments
with HCI. Thus PNC experiments in HCI can open a new field of research 
independent of the successes of PNC experiments in neutral atoms.

We calculate electroweak radiative corrections to the matrix 
element $F_{0}=\\ \langle ns_{1/2}|{\hat H_{\rm PNC}}|n'p_{1/2}\rangle$, 
which represents 
the basis for most parity nonconserving processes studied in atomic physics 
of highly charged ions. Here the operator ${\hat H}_{\rm PNC}$ indicates 
the parity nonconserving relativistic effective atomic Hamiltonian at the 
tree level. The electroweak radiative corrections will be provided by   
$\langle ns_{1/2}|{\hat H}_{\rm PNC}^{\rm rad}|n'p_{1/2}\rangle\,\equiv \,
F_{\rm SF}^{\rm rad}$.

Previously radiative electroweak corrections were calculated for the 
case of low momentum transfer ($q^{2}=0$) or for the low field case, 
that is valid for neutral atoms. We are following the work by Lynn and 
Sandars \cite {l6} who represented their results in the factorized form
\begin{equation}
{\hat H}_{\rm PNC}^{\rm rad}=
{\delta} {A_{\rm PNC}^{\rm rad}} \:{\hat H_{\rm PNC}}
\end{equation}
where the factor $\delta A_{\rm PNC}^{\rm rad}$ is independent of electron 
variables. This factorization is possible only in the low field case. 
Thus the quantity corresponding to $F_{\rm SF}^{\rm rad} $ in the low 
field case should be defined as 
\begin{equation}
F_{\rm LF}^{\rm rad}=\delta A_{\rm PNC}^{\rm rad}
\langle ns_{1/2}|\hat{H}_{\rm PNC}|n'p_{1/2}\rangle
\;.
\end{equation}
The deviation of the function $f^{\rm rad}\equiv 
\frac{F_{\rm SF}^{\rm rad}(Z)-F_{0}(Z)}{ F_{\rm LF}^{\rm rad}(Z)-F_{0}(Z)}$ 
from unity for large $Z$ values will manifest the existence of strong field 
effects for electroweak radiative corrections in the Standard Model.

The electroweak radiative corrections to the matrix element 
$\langle ns_{1/2}|{\hat H}_{\rm PNC}|n'p_{1/2}\rangle$ can be partitioned 
into the corrections 
$\langle ns_{1/2}|{\hat H}_{\rm PNC}^{\rm rad}|n'p_{1/2}\rangle$ to the PNC 
operator and to corrections to the wave functions. The corrections to the 
wave functions were not considered in \cite{l6}. Moreover, the full treatment 
requires the evaluation of radiative corrections to the total expression of 
the PNC atomic amplitude, including the PNC matrix element and the photon 
emission(absorption) matrix element. In this paper we will concentrate on 
radiative corrections to the PNC matrix element.

In section 2 of this paper we analyse the influence of strong fields 
on different electroweak corrections. In section 3 we evaluate electroweak 
radiative loop corrections to the PNC operator. In section 4 the loop 
corrections to wave functions are evaluated. Section 5 contains the 
discussion of the numerical results and conclusions.
  
\section{Analysis of electroweak radiative corrections in highly charged ions}
%\bigskip
In this paper we will consider only the nuclear spin-independent part of 
${\hat H}_{\rm PNC}$:
\begin{equation}
 {\hat H_{\rm PNC}}=A_{\rm PNC}\,{\gamma_{5}}{\rho_{N}(r)} \; ,
\end{equation}
where $\gamma_{5}$ is the Dirac matrix and $\rho_{N}(r)$ is the nuclear 
density. In the original Bouchiat formulation \cite{l7} the constant 
$A_{\rm PNC}$ reads 
\begin{equation}
A_{\rm PNC}=\frac{G_{F}Q_{W}}{2\sqrt{2}} \; ,
\end{equation}
where $G_{F}$ is the Fermi constant and $Q_{W}$ is the ``weak charge'' of 
the nucleus. At the tree level $Q_{W}$ is given by   
\begin{equation}
Q_{W}=-N+Z(1-4s^{2}) \; ,
\end{equation}
where $ s^{2}=\sin^{2}\theta_{W}$, $\theta_{W}$ is the Weinberg angle, $N$ 
is the number of neutrons in the nucleus.

In the following we will also utilize
the equivalent Sandars definition of $A_{\rm PNC}$ \cite{l18}:
\begin{equation}
{A_{\rm PNC}}=\frac
{\pi \alpha }
{4M_{Z}^{2}}
 P_{W} \; ,
\end{equation}
%\
where
\ $\alpha $ is the fine structure constant and $ M_{Z}$ is the mass of 
Z-boson. At the tree level
we have
\begin{equation}
P_{W}=\frac{-N+Z(1-4s^{2})}{s^{2}(1-s^{2})}\;.
\end{equation}
According to Lynn and Sandars \cite{l6}, all the electroweak radiative 
corrections can be divided in two classes. The first class, called 
``oblique'' corrections, corresponds to the Feynman graphs depicted in 
Fig. 1. These corrections can be incorporated into the running coupling 
constants, dependent on $q^{2}$. To include the ``oblique'' corrections 
into the PNC calculations for neutral atoms, Lynn and Sandars employ the 
running fine structure constant $\alpha^{*}(q^{2}=0)$, the sine of the 
Weinberg angle ${s^{*}}(q^{2}=0)$ and the mass of the Z-boson 
$ M^{2}_{Z}(q^{2}=0)$. The Fermi constant 
$G_{F}$ does not enter the Sandars description of PNC effects. In the case 
of atomic experiments $\alpha^{*}(q^{2}=0)=\alpha$, where $\alpha$ is the 
standard atomic value, $s$ and $M_{Z}^{*}$ are obtained from the LEP values 
\cite{l1} by scaling to $q^{2}=0$.

The remarkable feature of Sandars description is that for all heavy 
elements of experimental interest $P_{W}$ is very close to 
$-{\frac{16}{3}}N$ and is weakly dependent on $s^{2}$ \cite{l18}. Therefore 
it is convenient to introduce the quantity
\begin{equation}
{\tilde P}_{W}=-{\frac {3}{16N}} P_{W}
\;.
\end{equation}
Then the ''oblique'' radiative corrections can be included  in 
${\tilde {P}}_{W}$  
\begin{equation}
\tilde{P}_{W}={\tilde P}_{W}^{*}\,{(1+\delta_{P}^{M})} \; ,
\end{equation}
where ${\tilde P}_W^{*}$ is defined from Eq. (8) with 
$s^{2}=s^{* 2}(q^{2}=0)=0.2394$.  
The correction $\delta _{P}^{M}$ results as \cite{l6}
\begin{equation}
\delta _{P}^{M}=\frac {M_{Z}^{2}}
{[{M_{Z}^{*}}(q^{2}=0)]^{2}}-1 \simeq 0.0880
\;.
\end{equation}

Returning to our problem for HCI, we emphasize that we are interested to 
follow the change of the running constants in the interval from $q^{2}=0$ to 
$q^{2}=m^{2}_{e}=(0.5\ {\rm MeV})^{2}$. This interval is $10^{5}$ times 
smaller than the interval from $q^{2}=0$ to 
$q^{2}=M_{Z}^{2}=(91\ {\rm GeV})^{2}$. Therefore the value $\delta _{P}^{M}$ 
should be considered as field independent. Our goal is to search for the 
field dependent radiative corrections and to compare them with the value 
given by Eq. (11). In section 3 of this paper we will evaluate loop 
corrections directly, using the extension of the Furry picture of QED for 
tightly bound electrons.
In the Furry picture the electrons are considered from the beginning in 
the external field of the nucleus. The Feynman rules for QED in the Furry 
picture can be found, for example, in \cite{l19}.

To represent the basic atomic PNC matrix element in the Furry picture 
we have to consider first the Feynman graph corresponding to the exchange 
of a Z-boson between the atomic electron and the quark. Due to the vector 
current conservation the Z-boson coupling to the quarks in the case of an 
atom transforms to the Z-boson coupling to nucleons and to the nucleus. 
Thus the quark line in the Feynman graph should be replaced by the nuclear 
line.

Moreover, the interaction of the electron with the nucleus via the 
exchange of a Z-boson can be replaced by the interaction with the electroweak 
external field described by Eq. (4). This corresponds to the Feynman graph 
depicted in Fig. 2.

Now we can draw the Feynman graphs corresponding to loop corrections 
in the Furry picture of the Standard Model for tightly bound electrons. We 
suppose that the main contribution to the difference between the cases of 
$q^{2}=m^{2}_{e}$ (HCI) and $q^{2}=0$ (neutral atoms) arises from electron 
loops, since the heavier particle loops should be less sensitive to the 
strong field effect. In this respect 
we have to consider the graphs presented in 
Fig. 3 a)-d). In the evaluation of these graphs we utilize the Uehling 
approximation. Then the bound electron loop is approximated by the first 
term of the expansion in powers of the external potential (cf. Fig. 4 a)-d)). 
The Uehling approximation is justified even for tightly bound electrons in 
QED. The evaluation of electron loop corrections corresponding to Fig. 4 a), 
b) will be performed in section 3.
In section 4 we will investigate the corrections corresponding to 
Fig. 4 c), d), i.e. the corrections to the wave functions.

The other type of electroweak radiative corrections, called 
``specific'' \cite{l6}, corresponds to Feynman graphs displayed in Figs. 5 
and 6. Fig. 5 represents the contribution of the electron weak anapole 
moment, Fig. 6 corresponds to the vertices that describe the electromagnetic 
renormalization of the Z-boson coupling. According to our analysis only 
these graphs can contribute to the difference between the strong field and 
low field cases.

Lynn and Sandars \cite{l6} present the electroweak radiative
corrections to ${\tilde { P}}_{W}$ in the form:
\begin{eqnarray}
{\tilde { P}_{W}}&=&{\tilde { P}}_{W}^{*}(1+{\delta }^{M}_{P})+{\delta
}^{\rm anapole}_{P}
+{\delta }_{P}^{\rm vertex } 
\end{eqnarray}
where ${\tilde { P}_{W}^{*}}$ and $\delta^{M}_{P}$ are defined by 
Eqs. (10) and (11), $\delta_{P}$ are specific radiative corrections to 
$\tilde { P}_{W}$. In Eq. (12) we omitted small field-independent ''specific'' 
corrections given by the ''box'' Feynman graphs \cite{l6}.

 We present the results of our calculations in the form
\begin{equation}
\Delta_{\rm rad}=\Delta{\tilde {P}_{W}}+\delta_{P}^{\rm w.f.}
\end{equation}
with
\begin{eqnarray}
\Delta
{\tilde {P}_{W}}
&=&\delta_{P}^{\rm loop-op}
(f_{\rm loop-op}^{\rm rad}-1)\nonumber \\& &+\delta_{P}^{\rm anapole}
(f_{\rm anapole}^{\rm rad}-1)+
\delta_{P}^{\rm vertex}
(f_{\rm vertex}^{\rm rad}-1)
\;,
\end{eqnarray}
where 
$\Delta 
{\tilde P_{W}}
$ is the difference between the corrections for HCI and neutral atoms,
$\delta_{P}^{\rm loop-op.}$, $\delta_{P}^{\rm anapole}$ and 
$\delta_{P}^{\rm vertex }$ denote the $q^{2}=0$ limit for the different 
radiative corrections to the PNC operator, and
the functions $f^{\rm rad}$ are
defined in the Introduction. Actually in this paper only the term 
$\delta_{P}^{\rm loop-w.f.}$, the most important after $\delta_{P}^{M}$, 
will be calculated numerically. This term represents the loop corrections 
to the wave functions.

\section {Loop corrections to the PNC operator}
%\bigskip
We begin with the evaluation of the corrections of Fig. 4 a). To write down 
the corresponding matrix element we use the standard Feynman rules formulated
for the QED of tightly bound electrons \cite{l19}. These rules are
easily extended to the Standard Model calculations. The S-matrix element 
corresponding to the Feynman graph of Fig. 4 a) is given by
\begin{eqnarray}
S &=&(-i)^{2}g^{2}\int{d^{4}x_{1}d^{4}x_{2}d^{4}x_{3}}\,{\bar{\psi}}
_{ns_{1/2}}(x_{1}){\gamma^{\mu }(\gamma_{5}-\eta)}{\psi_{n'p_{1/2}}}(x_{1})
\nonumber \\
&& \times D_{\mu \nu}^{Z}(x_{1}-x_{2})
{\rm Tr}[{\gamma^{\nu}(\gamma_{5}-\eta)}S^{0}(x_{2}-x_{3})
{\gamma^{\lambda}}S^{0}(x_{3}-x_{2})]A_{\lambda}^{\rm ext}(x_{3}) \; ,
\end{eqnarray}
where $\eta =1-4\sin^{2}\theta_{W}$, Tr corresponds to the Dirac matrices,
that enter the electron loop. The Z-boson propagator $D^{Z}_{\mu \nu}$ in 
momentum space can be expressed as 
\begin{equation}
D^{Z}_{\mu \nu}(k)=-\frac{{4\pi}{i} \,g_{\mu \nu}}{k^{2}-M^{2}_{Z}+i0}
\;.
\end{equation}
We will use the expression for $D^{Z}_{\mu \nu}(x_{1}-x_{2})$ in coordinate 
space 
\begin{equation} 
D^{Z}_{\mu \nu }(x_{1}-x_{2})=
\int{\frac{d\omega}{2\pi}}
\:
e^{-i \omega(t_{1}-t_{2})}
{D^{Z}_{\mu \nu}}
(\vec x_{1}-\vec x_{2},\omega)
\end{equation}
where
\begin{equation}
D^{Z}_{\mu \nu }({\vec x_{1}}-{\vec x_{2}},\omega )=-i 4\pi g_{\mu \nu }
\int{\frac{d^{3} k}{2\pi}}
\frac
{
e^{i {\vec k}\ ({\vec x}_{1}-{\vec x}_{2})}
}
{\omega^{2}-{\vec k}^{2}-M_{Z}^{2}+i 0}
\;.
\end{equation}
The external electromagnetic field $A_{\nu}^{\rm ext}(x)$ reads 
\begin{equation}
A^{\rm ext}_{\nu}(x)=
{\delta_{\nu 0}}\,eU(\vec{x})
\end{equation}
where $U(\vec {x})$ is the electric field of the nucleus (point-like or
extended). $S^{0}(x-y)$ is the free electron
propagator
\begin{equation}
S^{0}(x-y)=
{\frac{1}{(2\pi)^{4}}}
\int d^{4}p\, S^{0}(p) \,e^{-i p(x-y)}
\end{equation}
with
\begin{equation}
S^{0}(p)=i {\frac{p\!\!\!/+m_{e}}{p^{2}-m^{2}_{e}}}\;.
\end{equation}

The wave functions ${\bar \psi }_{ns_{1/2}},
{\psi_{n'p_{1/2}}}$ are the eigenvectors of the Dirac equation for the 
electron in the field of the nucleus
\begin{equation}
\psi_{n}(x)= e^{-iE_{n}t}
\,
\psi_{n} (\vec x) \; ,
\end{equation}
where $E_{n}$ are the corresponding Dirac eigenvalues.
\ The Standard Model constant $g$ is equal to
\begin{equation}
g^{2}=\frac {e^{2}}{s^{2}} \; ,
\end{equation}
where $e$ is the electron charge. We employ the pseudoeuclidean metric
with the usual metric tensor $g_{\mu \nu }$. $\gamma_{\mu }$, $\gamma_{5}$
are the usual Dirac matrices.

The $S$-matrix element is connected with the amplitude by the relation
\begin{equation}
S_{if}={2\pi}i
\,
\delta (E_{i}-E_{f})
\,
M_{if}
\end{equation}
where $M_{if}$ is the amplitude and $E_{i}$, $E_{f}$ 
are the initial and final state energies of the system. Transforming to the 
momentum space in expression (15) and integrating over the frequency 
variables, we obtain
\begin{equation}
M_{if}=-{\frac {(1-4s^{2})}{16c^{2}s^{2}}}
{\frac{1}{8\pi^{2}}}
\int d^3p_{1} d^3p_{2}\,
{\bar \psi_{ns_{1/2}}}
(\vec p_{1})\,
{
\frac {eU(\vec q\,)}
{\vec q^{\,2}+M_{Z}^{2}}
}\,
\gamma_{0}
\gamma_{5}\,
\Pi
(0,{\vec q}^{\,2})
{\psi_{n'p_{1/2}}}
(\vec p_{2})
\end{equation}
where ${\vec q}={\vec p}_{1}-{\vec p}_{2}$.
In the expression (25) we retain only the parity violation terms. These 
terms contain the $\gamma_{5}$
matrix in one of the vertices connected with Z-boson. The vertex connected 
with the loop yields a zero result due to the identities:
\begin{eqnarray}
{\rm Tr}
(\gamma_{\mu }
 \gamma_{\nu } 
 \gamma_{5})&=&0\;,\\
{\rm Tr}
(\gamma_{\mu }
 \gamma_{\nu }
 \gamma_{\alpha }
 \gamma_{\beta }
 \gamma_{5})&=&
 i 
 \epsilon_{\mu \nu \alpha \beta }
\;,
\end{eqnarray}
where 
$\epsilon_{\mu \nu \alpha \beta }$
is the unit antisymmetrical tensor. This tensor appears in the combination 
with the symmetrical product
$p_{\mu }p_{\nu }$, so that
\begin{equation}
\epsilon_{\mu \nu \alpha \beta } \,p_{\mu } p_{\nu }=0\; .
\end{equation}

The function 
$\Pi
({\vec q}^{\:2})$
is divergent and should be renormalized. We shall use from the very 
beginning the known renormalized expression $\Pi_{R}(q^{2})$
(cf. for example \cite{l20})
\begin{equation}
\Pi_{R}
(q^{2})
={
\frac{2i e^{2}}
     {2\pi^{2}}
}
q^{2}
\int\limits_{0}^{1}
dx
\:
x(1-x)
\ln 
\left[1-\frac{q^{2}}{m^{2}_{e}}
x(1-x)\right]
\;.
\end{equation}
Since all the integrations in Eq. (25) after the insertion of the 
renormalized expression (29) are convergent, we can omit ${\vec q}^{\:2}$ 
in the denominator. In the case of the pure Coulomb potential with 
\begin{equation}
U(\vec q\,)=4\pi 
\frac {eZ}{{\vec q}^{\:2}}
\end{equation} 
we obtain
\begin{eqnarray}
M_{if}
&=&
{\frac
{(1-4s^{2})}
{2c^{2}s^{2}}
}
{\frac
{e^{3}Z}
{(2\pi )^{2}}
}
\frac{1}
{M^{2}_{Z}}
\int\limits_{0}^{1}
dx
\:
x(1-x)
\nonumber \\
& &
\int d^3p_{1} d^3p_{2}\,
{\bar \psi}_{ns_{1/2}}
(\vec p_{1})
\gamma_{0}
\gamma_{5}
\ln 
\left[1+\frac{{\vec q}^{\:2}}{m^{2}_{e}}
x(1-x)\right]
\psi_{n'p_{1/2}}
(\vec p_{2})
\;.
\end{eqnarray}
We consider first the low field limit of Eq. (31). Then 
${\vec q}^{\:2}/m^{2}_{e}\ll 1$ and we can write
\begin{eqnarray}
M_{if}&=&-
{\frac
{(1-4s^{2})}
{16c^{2}s^{2}}
}
{\frac
{e^{2}Z}
{\pi^{3}}
}
\frac{e^{2}}
{m^{2}_{e}M^{2}_{Z}}
\int \limits_{0}^{1}dx
\:
{x^{2}(1-x)^{2}} 
\nonumber\\ 
& &
\times\left[
\int  d^3p_{1} d^3p_{2}\,
{\varphi }_{ns_{1/2}}^{+}
({\vec p}_{1})
{{\vec q}^{\:2}}
{\chi }_{n'p_{1/2}}
({\vec p}_{2})
+
\int d^3p_{1} d^3p_{2}\, 
{\chi }_{ns_{1/2}}^{+}
({\vec p}_{1})
{{\vec q}^{\:2}}
{\varphi }_{n'p_{1/2}}
({\vec p}_{2})
\right]
\,,
\end{eqnarray}
where $\varphi ,\chi $ are the upper and lower components of the Dirac 
wave function, respectively.  

Transforming to the coordinate representation we obtain
\begin{eqnarray}
M_{if}&=&
{\frac
{(1-4s^{2})}
{16c^{2}s^{2}}
}
{\frac
{e^{2}Z}
{\pi^{3}}
}
\frac{e^{2}(2\pi )^{3}}
{m^{2}_{e}M^{2}_{Z}}
\int \limits_{0}^{1}dx
\:
{x^{2}(1-x)^{2}} 
\nonumber\\ 
& &
\times\left[
\int d^3r_{1} d^3r_{2}\,  
{\varphi }_{ns_{1/2}}^{+}
({\vec r}_{1})
(\vec\nabla_{1}+\vec\nabla_{2})^{2}
\delta ({\vec r}_{1})
\delta ({\vec r}_{2})
{\chi }_{n'p_{1/2}}
({\vec r}_{2})
\right.
\nonumber\\
& &
+
\left.
\int d^3r_{1} d^3r_{2}\, 
{\chi }_{ns_{1/2}}^{+}
({\vec r}_{1})
(\vec \nabla_{1}+\vec\nabla_{2})^{2}
\delta ({\vec r}_{1})
\delta ({\vec r}_{2})
{\varphi }_{n'p_{1/2}}
({\vec r}_{2})
\right]
\;.
\end{eqnarray}
Finally we find:
\begin{eqnarray}
M_{if}& =&
\frac
{1-4s^{2}}
{12c^{2}s^{2}}
\frac
{e^{4}Z}{\pi }
\frac{\pi }
{m^{2}M^{2}_{Z}}
\int \limits_{0}^{1}dx
\;
x^{2}(1-x)^{2}  
\left[
%(
\nabla^{2}
{\varphi }_
{ns_{1/2}}^
{+}
(0)
%)
{\chi }_
{n'p_{1/2}}
(0)
\right. \nonumber\\
& &
+
{\varphi }_
{ns_{1/2}}^
{+}
(0)
\nabla^
{2}
{\chi }_{n'p_{1/2}}
(0)+2
\vec\nabla
{\varphi }_
{ns_{1/2}}^
{+}
(0)
\cdot \vec\nabla
{\chi }_{n'p_{1/2}}
(0)
\nonumber\\
& &
+\left. 
{\chi }_{n'p_{1/2}}
^{+}
(0)
\nabla^{2}
{\varphi }_
{n'p_{1/2}}
(0)
+2
\vec\nabla
{\chi }_
{ns_{1/2}}^
{+}
(0)
\cdot \vec\nabla
{\varphi }_{n'p_{1/2}}
(0)
\right]
\;.
\end{eqnarray} 
Now we compare Eq. (34) with the matrix element of ${\hat H}_{\rm PNC}$ 
given by Eq. (4). We can write the latter in the form
\begin{equation}
({\hat H}_{\rm PNC})_{if}=
{
\frac
{\pi \alpha }
{4M_{Z}^{2}}
}
P_{W}
\varphi_{ns_{1/2}}^{+}(0)\,\chi_{n'p_{1/2}}(0)\; .
\end{equation}
This comparison leads to the estimate
\begin{equation}
\delta_{P}^{\rm loop-op}\approx
\frac{1}{15\pi }
\left({-\frac
{Z}
{N}}\right)
(1-4s^{2})
\alpha
(\alpha Z)^{2}
\;.
\end{equation}
Now we evaluate the correction of Fig. 4b. The corresponding matrix 
element reads
\begin{equation}
S =
(-\imath ) e^2
\int
d^{4}x_{1}d^{4}x_{2}d^{4}x_{3}
\,
\bar \psi_{ns1/2}
(x_{1})
\gamma^{\mu }
\psi_{n'p1/2}
(x_{1})
D_{\mu \nu }^{\gamma }
(x_{1}-x_{2})
\Pi_{R}(x_{2},x_{3})
Z^{\nu  }
(x_{3})
\end{equation}
where $D^{\gamma }_{\mu \nu }(x_{1}-x_{2})$ 
is the photon propagator in Feynman gauge
\begin{equation}
D^{\gamma }_{\mu \nu }
(x_{1}-x_{2})=
\int{\frac{d\omega}{2\pi}}
\:
e^{-i \omega(t_{1}-t_{2})}
{D^{\gamma }_{\mu \nu}}
(\vec x_{1}-\vec x_{2},\omega)
\end{equation}
with
\begin{equation}
D_{\mu \nu }^{\gamma }
(\vec x_{1}-\vec x_{2},\omega)=
-i 4\pi \,g_{\mu\nu}
\int
{\frac
{d^{3}k}
{(2\pi )^{3}}
}
\frac
{e^{i\vec k({\vec x}_{2}-{\vec x}_{1})}}
{\omega^{2}-{\vec k}^{2}+i \epsilon}
\;,
\end{equation}
and 
$Z_{\nu }$ 
is the external electroweak field defined by Eq. (4)
\begin{equation}
Z_{\nu }^{e}=\delta_{\nu 0}
{\rm A}_{\rm PNC}
\gamma_{5}
\,\rho_{N}
(\vec x\,)
\;.
\end{equation}

It turns out that the matrix element (37) is exactly zero. Returning from Fig. 4 b) to Fig. 3 b) we expand the bound electron loop
in powers of the external potential (19). This expansion will contain an
increasing number of Dirac matrices $\gamma^\alpha$ together with one
matrix $\gamma_5$ from Eq. (40).
The trace of the product of an arbitrary number of Dirac matrices can 
be reduced to traces of lower products. 
Then, using the Eqs. (26), (27) we will obtain a zero result for an 
arbitrary term of the bound electron 
loop expansion. Thus, the correction of Fig. 3 b) is absent for the nuclear spin-independent part of 
$\hat H_{\rm PNC}$.

\section{Loop corrections to wave functions}
%\bigskip
We start with the investigation of the graphs Fig. 4 c), d).
 Unlike the graphs 4 a), b) these graphs are reducible \cite{l19}. This 
 means that the initial state of the
system (the ``reference'' state) can be found among the intermediate states. 
The presence of the reference state in the sums over intermediate states 
leads to singularities that have to be avoided. For the solution of the 
``reference state'' problem for the diagonal matrix element (i.e. the energy 
correction) the adiabatic approach of Gell-Mann and Low \cite{l21}, modified 
by Sucher \cite{l22} is used most frequently \cite{l19}. The extention of 
this approach to the nondiagonal matrix element can be most naturally 
formulated within the framework of the line profile QED theory \cite{l23}.

Actually the graphs in Fig. 4 do not correspond to any amplitude, since the
amplitude should describe some process in an atom. Still the graphs 
Fig. 4 a), b) are
irreducible and Eq. (24) formally can be applied to them as well.

\
\ In the case of the graphs 4 c), d) we have to remember that the PNC matrix 
element enters necessarily in some complex amplitude describing the atomic 
process.
In the simplest case it can be the process of photon emission by an atomic
electron in one-electron ions. We will consider the situation when only 
two levels of opposite parity $ns_{1/2}$ and $n'p_{1/2}$
are mixed by the electroweak interaction. Actually this situation does not 
occur in one-electron HCI, but can be found in two-electron ions 
\cite{l19}, \cite{l24}, \cite{l25}.

\
\ Instead of the graph of Fig. 4 c) we have now to consider the graph in 
Fig. 7 a). The corresponding S-matrix element is given by 
\begin{eqnarray}
S =
(-i )^{3}
e^{2}
\int
d^{4}x_{1}
\,
d^{4}x_{2}
\,
d^{4}x_{3}
\,
{\bar \psi }_{n''s_{1/2}}(x_{1})
\gamma^{\mu }
{\rm A}_{\mu }^{(\omega )*}(x_{1})
\nonumber \\
 S_{e}
(x_{1},x_{2})
\gamma^{\nu }
{\rm A}^{{\rm ext}}_{\nu }
(x_{2})
S_{e}
(x_{2},x_{3})
\gamma^{\lambda }
Z_{\lambda }^{(e)}
(x_{3})
\psi_{ns_{1/2}}(x_{3})
\;,
\end{eqnarray}
where 
${\rm A}^{(\omega )*}_{\mu }
(x)$
is the wave function of the emitted photon 
\begin{equation}
{\rm A}_{\mu }^{(\omega )*}(x)=
\sqrt 
{
\frac
{4\pi }
{2V\omega }
}
{\vec e}^{\ (\omega )*}
\,
e^{-i (\omega t-\vec k \vec x)}
\;.
\end{equation}  
$\vec e$ is the polarization vector, $\omega$, $\vec k$ are the frequency
and the wave vector of the photon. The external electromagnetic potential 
${\rm A}_{\nu }^{{\rm ext}}(x)$ is
\begin{equation}
{\rm A}^{{\rm ext}}_{\nu }
(x)=
\delta_{\nu 0}
{\rm V}_{U}
(x)
\;,
\end{equation}
where ${\rm V}_{U}(x)$ is the Uehling potential \cite{l26}, \cite{l27}
\begin{equation}
{\rm V}_{U}(x)=
{
\frac 
{2e^{3}Z}
{3\pi x}
}
\int^{\infty }_{1}
e^{-2mxy}
\left(
1+
\frac{1}{2 y^{2}}
\right)
\frac
{\sqrt{y^{2}-1}}
{y^{2}}
\,
dy
\;.
\end{equation}
$S_{e}(x_{1},x_{2})$ denotes the electron propagator in the external
field \cite{l27}
\begin{equation}
S_{e}(x_{1},x_{2})=
{\frac {1}
{2\pi i} 
}
\int d\omega '
\;
e^{i \omega '(t_{1}-t_{2})}
\sum_{m}
{\frac
{
\psi_{m}(\vec x_{1})
{\bar \psi}_{m}
(\vec x_{2})
}
{
E_{m}
(1-i 0)
+\omega^{,}
}}\; .
\end{equation}
\ The sum over $m$ in Eq. (45) is extended over the complete spectrum of 
the Dirac
Hamiltonian for the electron in the field of the nucleus.

\
The integration over the time variables in Eq. (41) with the help of 
formula (24) yields
\
\begin{equation}
{\rm M}_{n''s_{1/2};ns_{1/2}}=
-\imath e^{2}
\sum_{m',m''}
\frac { 
\langle 
n''s_{1/2}|
{\vec e}
{\vec {\rm A}^{(\omega )*}}|
m'
\rangle
\langle
 m'|
{\rm V}_{U}|
m''
\rangle 
\langle 
m''|
H_{\rm PNC}|
ns_{1/2}
\rangle 
}
{
(E_{m'}-E_{ns_{1/2}})(E_{m''}-E_{ns_{1/2}})}
\,.
\end{equation}
\ There are singular terms in the sums over $m',m''$ when $m',m''=ns_{1/2}$.
These singularities can be avoided by the use of the line profile theory
\cite{l23}. 
Actually in the framework of this theory the singular terms should
be omitted, but some additional terms containing derivatives of the
potentials with respect to the energy can arise. In our case, due to the
independence of the potentials ${\rm V}_{U},H_{\rm PNC}$ on the energy, 
these additional terms are absent.

\ Remembering now that we assumed that only one level of opposite parity
$n'p_{1/2}$ is close to the initial level $ns_{1/2}$, we set $m'=n'p_{1/2}$.
Then
\begin{equation}
{\rm M}_{n''s_{1/2};ns_{1/2}}=
 e^{2}
\frac
{
\langle n''s_{1/2}|
\vec e
\vec {\rm A}^{(\omega )*}|
n'p_{1/2}\rangle
}
{
E_{n'p_{1/2}}-E_{ns_{1/2}}
}
\sum_{
m\ne 
ns_{1/2}}
\frac
{
\langle n'p_{1/2}|
{\rm V}_{U}|
m\rangle
\langle m|
H_{\rm PNC}|
ns_{1/2}\rangle
}
{
E_{m}-E_{ns_{1/2}}
} \; .
\end{equation}
\ Performing the same calculations for the graph in Fig. 4d) (i.e. refering 
to the graph Fig. 8b)) and using the same
assumptions we would obtain the expression 
\begin{equation}
{\rm M}_{n''s_{1/2};ns_{1/2}}= e^{2}
\frac
{
\langle n''s_{1/2}|
\vec e
\vec {\rm A}^{(\omega )*}|
n'p_{1/2}\rangle
}
{
E_{n'p_{1/2}}-E_{ns_{1/2}}
}
\sum_{
m\ne 
ns_{1/2}}
\frac
{
\langle n'p_{1/2}|
H_{\rm PNC}|
m\rangle
\langle m|{\rm V}_{U}|
ns_{1/2}\rangle
}
{
E_{m}-E_{ns_{1/2}}
}\; .
\end{equation}
\ The parts of the expressions (47) and (48) containing the sums over $m$ 
yield evidently the corrections to the wave functions in the PNC matrix 
element.
\  
\ Note that we can consider the graph in Fig. 7a as a radiative loop
correction to the photon emission matrix element in the parity violating 
photon emission amplitude.

\
\ Moreover, apart from the graphs in Fig. 7 we have, in principle, to 
consider
also the graph in Fig. 8 which describes exclusively the radiative 
corrections to the photon
emission. However, in this paper we will restrict ourselves only to the
corrections to the PNC matrix element.
\
\ Using Eq. (41) and remembering that in the low field limit the energy
denominators in Eqs. (47) and (48) are of the order $m(\alpha Z)^{2}$, we
obtain the estimate
\begin{equation}
\delta_{P}^{\rm loop-w.f.}\approx \alpha (\alpha Z)^{2}
\;.
\end{equation} 
Comparing (49) with the estimate (36) we conclude, that the corrections to 
the wave functions are dominant.

%\bigskip
\section{Numerical results}
%\bigskip
\ In this paper we provide numerical results only for the leading loop 
corrections
corresponding to the Feynman graphs in Fig. 4 c), d).

\
\ These leading corrections are the corrections to wave functions
discussed in section 4. The loop correction corresponding to Fig. 4 a) is
suppressed by the factor $(1-4s^{2})$ in Eq. (25) and the correction
corresponding to Fig. 4b is absent for the nuclear spin-independent 
$\hat H_{\rm PNC}$. The electron anapole moment correction is suppressed 
again by the
factor $(1-4s^{2})$. The vertex corrections do not contain this suppression 
and
should be compiled together with the loop corrections of Fig. 4 c), d). 
The vertex corrections will be treated separately in a subsequent paper.
\
\ Thus, we retain here only the last term in Eq. (13). We performed the
calculations for the PNC matrix element including the Uehling potential in 
the Dirac equation for the atomic electrons. This equation was solved 
numerically with the computer code published in \cite{l28}. Then we 
subtracted the same matrix element calculated without the Uehling 
potential.
\
\ The results are listed in Table 1. The value of $A_M$ given in the second 
column is: $A_M =$ Atomic Weight. 
%${-}\frac{Z}{m_{p}}$, where $m_{p}$ is proton mass. 
In the third column the nuclear radius is given. The fourth 
column in this Table
represents the values for the PNC matrix element without the loop
correction, the fifth column is the same matrix element calculated with
the
Uehling potential taken into account. 
\
\ 
An extended nucleus with an uniform charge distribution is employed 
throughout all the calculations. The nuclear radius $R$ is taken to be
\begin{equation}
R=1.2\,A_M^{1/3}\;{\rm fm}
\;.
\end{equation}
\ In Table 2 we present the values of $\delta_{P}^{\rm loop-w.f.}$ for 
different $Z$ values. For $Z=92$ this correction is 7 times smaller than 
main correction (11) that is insensitive to the strong field. From QED 
calculations we know, that the vacuum polarization corrections are strongly 
sensitive to the field, i.e., results obtained in the low field approximation 
and extrapolated to the strong field case differ from the accurate strong 
field calculation by 100\% and more.
\
\ From Tables 1, 2 we can deduce, that the strong field effect for the 
electroweak radiative corrections in HCI exceeds 10\symbol{"25} for $Z=92$. 
This is an order of magnitude higher than could be expected from the simple 
extrapolation of the Lynn and Sandars \cite{l6} values.
The results obtained here demonstrate that the experiments with
HCI would provide the possibility to test the  Standard Model in the strong
field.
\
\ The most likely candidate for future PNC experiments with HCI is the
He-like uranium ion \cite{l24}, \cite{l25}, \cite{l29}. The theory developed 
in the present paper allows
for the evaluation of all the electroweak radiative corrections for this ion 
as well. 

\section{Acknowledgments}
%\bigskip
\looseness=-1
\  The authors are grateful to Dr. M. G. Kozlov for fruitful discussions. 
The work of I. B. and L. L. was supported by the Russian Fund for
Fundamental Investigations, grant $N\stackrel{0}{=}99-02-18526$, the work 
of I. B. also by the Administration of St. Petersburg, grant $N\stackrel{0}{=}M-97-2.2d-974$. G.S. acknowledges support by the BMBF, the DFG, and by GSI (Darmstadt).

\newpage
\begin{table}
\begin{center}
%\begin{tabular}{|c|c|c|c|c|}
\begin{tabular}{rrrrr} 
%\hline
 Z &    $A_{M}$    & $R_{\rm nucl}$ &     PNC       &  PNC(Uehl.)   \\
\hline
1  &  1.007    & 1.212  & .1954019E-17  & .1954049E-17  \\
2  &  4.001    & 1.921  & -.7939763E-15 & -.7940027E-15 \\    
3  &  6.939    & 2.307  & -.8095202E-14 & -.8095641E-14 \\
4  &  9.010    & 2.517  & -.3254761E-13 & -.3255015E-13 \\
5  &  10.807   & 2.675  & -.9200810E-13 & -.9201771E-13 \\
6  &  12.007   & 2.770  & -.1960428E-12 & -.1960690E-12 \\
7  &  14.002   & 2.916  & -.4258223E-12 & -.4258927E-12 \\
8  &  15.995   & 3.048  & -.8348039E-12 & -.8349702E-12 \\
9  &  18.994   & 3.228  & -.1698600E-11 & -.1698998E-11 \\     
10 &  20.173   & 3.293  & -.2637995E-11 & -.2638717E-11 \\
20 &  40.069   & 4.140  & -.9367160E-10 & -.9374468E-10 \\
30 &  65.363   & 4.874  & -.1014421E-08 & -.1015930E-08 \\
40 &  91.198   & 5.446  & -.5932577E-08 & -.5946879E-08 \\
50 &  118.662  & 5.945  & -.2623394E-07 & -.2632746E-08 \\
60 &  144.207  & 6.345  & -.9515140E-07 & -.9562729E-07 \\
70 &  173.001  & 6.742  & -.3259331E-06 & -.3281342E-06 \\
80 &  200.546  & 7.082  & -.1044648E-05 & -.1053974E-05 \\
82 &  207.155  & 7.159  & -.1325790E-05 & -.1338273E-05 \\
90 &  231.989  & 7.434  & -.3371690E-05 & -.3410852E-05 \\
92 &  234.993  & 7.466  & -.4167152E-05 & -.4218238E-05 \\
92 &  238.000  & 7.498  & -.4249630E-05 & -.4301619E-05 \\
%\hline
\end{tabular}
\end{center}
\caption{
\label{tab1}
Loop correction to the wave functions in the PNC matrix 
element for $n=n'=2$. The values of the matrix elements are given in eV. 
The nuclear radius is given in units of fm. }
\end{table}

\newpage
\begin{table}
\begin{center}
%\begin{tabular}{|c|c|c|c|}
\begin{tabular}{rrrr}
%\hline
 Z &    $A_{M}$    & $R_{\rm nucl}$ &  $\delta_{\rm P}^{\rm loop  w.f.}$  \\
\hline
1  &  1.007    & 1.212  & .1528E-04 \\
2  &  4.001    & 1.921  & .3332E-04 \\    
3  &  6.939    & 2.307  & .5414E-04 \\
4  &  9.010    & 2.517  & .7790E-04 \\
5  &  10.807   & 2.675  & .1044E-03 \\
6  &  12.007   & 2.770  & .1339E-03 \\
7  &  14.002   & 2.916  & .1653E-03 \\
8  &  15.995   & 3.048  & .1992E-03 \\
9  &  18.994   & 3.228  & .2343E-03 \\     
10 &  20.173   & 3.293  & .2736E-03 \\
20 &  40.069   & 4.140  & .7801E-03 \\
30 &  65.363   & 4.874  & .1488E-02 \\
40 &  91.198   & 5.446  & .2410E-02 \\
50 &  118.662  & 5.945  & .3564E-02 \\
60 &  144.207  & 6.345  & .5001E-02 \\
70 &  173.001  & 6.742  & .6753E-02 \\
80 &  200.546  & 7.082  & .8927E-02 \\
82 &  207.155  & 7.159  & .9415E-02 \\
90 &  231.989  & 7.434  & .1161E-01 \\
92 &  234.993  & 7.466  & .1223E-01 \\
92 &  238.000  & 7.498  & .1225E-01 \\
%\hline
\end{tabular}
%\
\end{center}
%\
%\begin{center}
%\  Table 2. The wave function contribution to $\Delta_{\rm rad}$ (Eq. (13)).
%\end{center}
\caption{\label{tab2}
The wave function contribution to $\Delta_{\rm rad}$ (Eq. (13)).
}
\end{table}

\
\newpage
\begin{figure}
\begin{center}
\begin{picture}(460,320)(0,0)
\Line(0,40)(0,220)
\Line(120,40)(120,220)
\GCirc(60,130){30}{0.5}
\Photon(0,130)(30,130){5}{3}
\Photon(90,130)(120,130){5}{3}
\Text(150,130)[]{\large $+$}
\Line(180,40)(180,220)
\Line(360,40)(360,220)
\GCirc(230,130){30}{0.5}
\GCirc(310,130){30}{0.5}
\Photon(180,130)(200,130){5}{3}
\Photon(260,130)(280,130){5}{3}
\Photon(340,130)(360,130){5}{3}
\Text(400,130)[]{\large $+\,\,\,\,\ldots $}
\Text(60,30)[]{a)}
\Text(270,30)[]{b)}
\end{picture}
\end{center}
\
\caption{ \label{fig1} 
The Feynman graphs corresponding to the ``oblique'' corrections.
The solid line correspond to fermions, the wavy line correspond to vectors 
($W^{+},W^{-},Z,\gamma $), the dark circles denote the fermion, vector and 
scalar loops.}%
\end{figure}
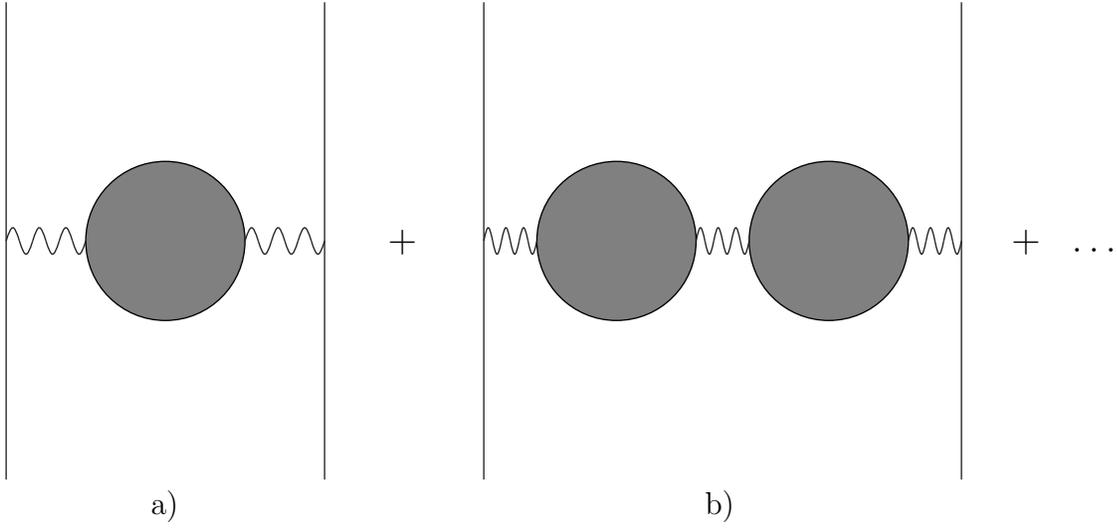

\newpage
\begin{figure}
\begin{center}
\begin{picture}(460,400)(0,0)
\Line(130,30)(130,300)
\Line(140,30)(140,300)
\Photon(135,165)(220,165){5}{8}
\Line(215,170)(225,160)
\Line(215,160)(225,170)
\Text(185,180)[b]{$Z$}
\Text(135,315)[b]{$n'p_{1/2}$}
\Text(135,20)[t]{$ns_{1/2}$}
\Vertex(135,165){5}
\end{picture}
\end{center}
\
\caption{ \label{fig2} The Feynman graph corresponding to the basic 
atomic matrix element in the Furry picture. 
The double solid line denotes the electron in the field of the
nucleus. The wavy $Z$ line with the cross at the end denotes the interaction
with the field given by Eq. (4).}
%it's the end of Fig.3
\end{figure}
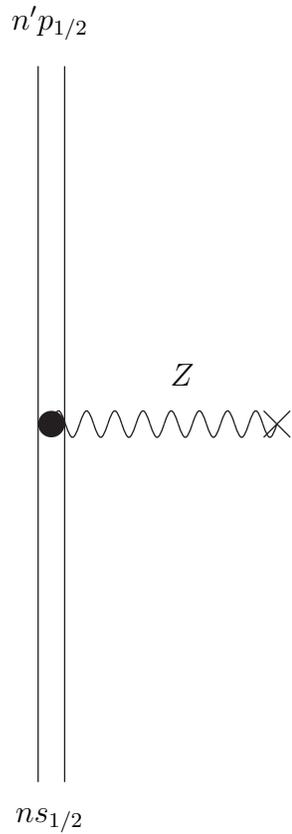

\newpage
\begin{figure}
\begin{center}
\begin{picture}(460,222)(0,0)
\Line(40,40)(40,200)
\Line(50,40)(50,200)
\Photon(50,120)(100,120){5}{5}
\Vertex(45,120){5}
\Vertex(105,120){5}
\BCirc(130,120){30}
\BCirc(130,120){20}
\Text(75,135)[b]{$Z$}
\Text(45,35)[t]{$ns_{1/2}$}
\Text(45,210)[b]{$n'p_{1/2}$}
\Text(45,0)[b]{a)}
\Vertex(105,120){5}
			      %Fig.4b)vniz

\Line(300,40)(300,200)
\Line(310,40)(310,200)
\put(345,120){circle*{5}}
\Photon(310,120)(345,120){5}{3}
\Vertex(305,120){5}
\Vertex(345,120){5}
\Vertex(395,120){5}
\BCirc(370,120){30}
\BCirc(370,120){20}
\Vertex(345,120){5}
\Photon(400,120)(430,120){5}{3}
\Line(425,125)(435,115)
\Line(425,115)(435,125)
\Text(415,135)[b]{$Z$}
\Text(305,35)[t]{$ns_{1/2}$}
\Text(305,210)[b]{$n'p_{1/2}$}
\Text(325,135)[b]{$\gamma $}
\Text(305,0)[b]{b)}
\Vertex(395,120){5}
\end{picture}
\end{center}
			      %Fig.4c vniz                               
\begin{center}
\begin{picture}(460,240)(0,0)
\Line(40,40)(40,200)
\Line(50,40)(50,200)
\Photon(50,90)(130,90){5}{8}
\Photon(50,140)(100,140){5}{5}
\BCirc(130,140){30}
\BCirc(130,140){20}
\Vertex(105,140){5}
\Vertex(45,140){5}
\Vertex(45,90){5}
\Vertex(105,140){5}
\Line(135,95)(125,85)
\Line(125,95)(135,85)
\Text(75,105)[b]{$Z$}
\Text(75,155)[b]{$\gamma $}
\Text(45,35)[t]{$ns_{1/2}$}
\Text(45,225)[b]{$n'p_{1/2}$}
\Text(45,0)[b]{c)}              
\Line(300,40)(300,200)
\Line(310,40)(310,200)
\Photon(310,90)(340,90){5}{3}
\BCirc(370,90){30}
\BCirc(370,90){20}
\Vertex(345,90){5}
\Vertex(305,140){5}
\Vertex(305,90){5}
\Photon(310,140)(370,140){5}{6}
\Line(365,145)(375,135)
\Line(365,135)(375,145)
\Text(340,155)[b]{$Z$}
\Text(305,35)[t]{$ns_{1/2}$}
\Text(305,225)[b]{$n'p_{1/2}$}
\Text(325,105)[b]{$\gamma $}
\Text(305,0)[b]{d)}
\end{picture}
\end{center}
\caption{
\label{fig3}
The Feynman graphs with electron loops contributing to the electroweak 
radiative corrections. Notations are the same as in Fig. 2. The wavy $Z$ 
line denotes the Z-boson, the wavy $\gamma $ line denotes the photon.}
\end{figure}
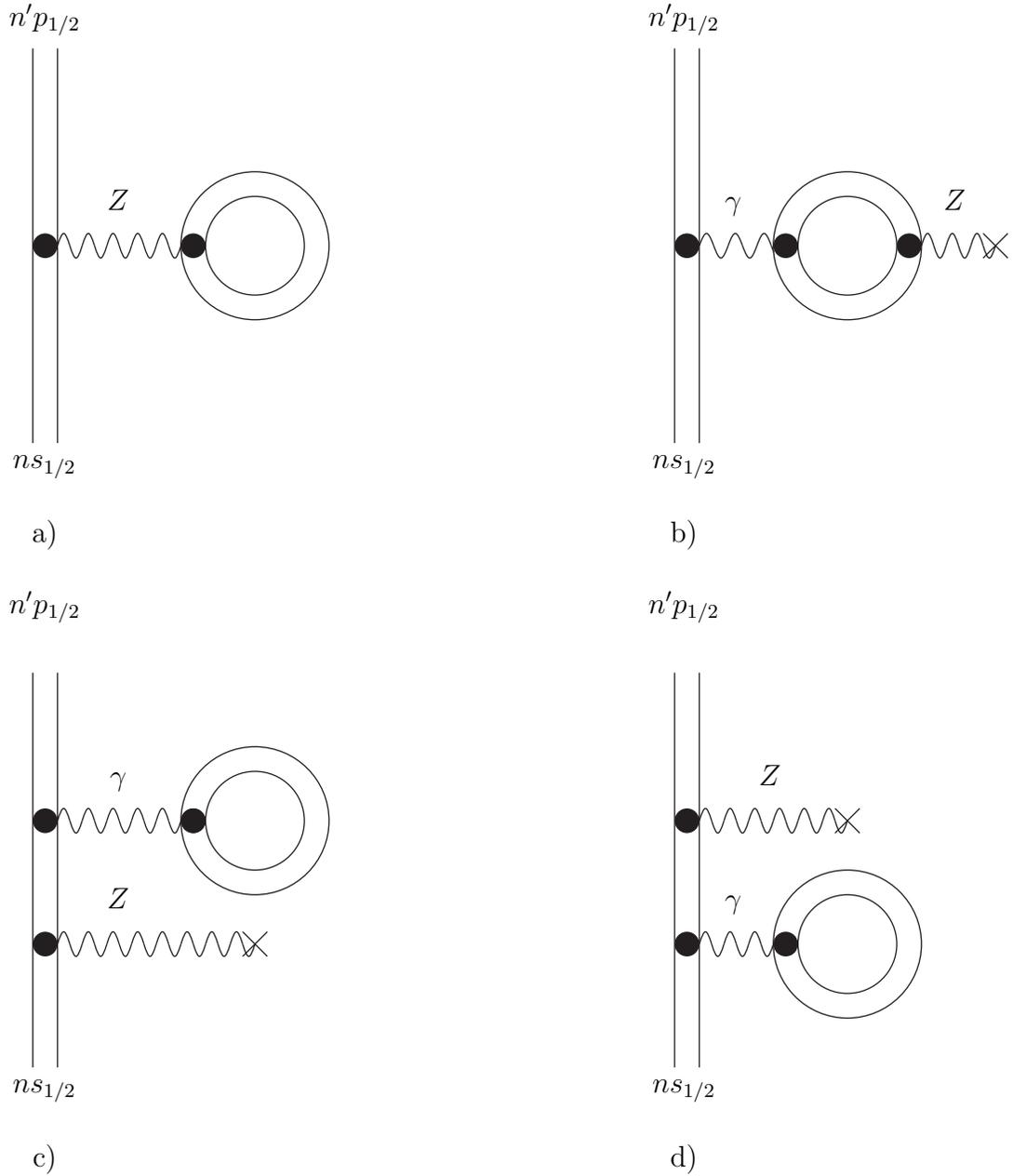

\newpage
\begin{figure}
\begin{center}
\begin{picture}(460,240)(0,0)
			     %Fig.5a vniz
\Line(60,60)(60,220)
\Line(70,60)(70,220)
\Photon(60,140)(100,140){5}{4}
\Photon(140,140)(180,140){5}{4}

\BCirc(120,140){20}
\Text(80,155)[b]{$Z$}
\Text(160,155)[b]{$\gamma $}
\Text(65,55)[t]{$ns_{1/2}$}
\Text(65,225)[b]{$n'p_{1/2}$}
\Text(65,5)[b]{a)}
\Line(175,145)(185,135)
\Line(175,135)(185,145)
\Vertex(65,140){5}
			     %Fig.5b vniz
\Line(280,60)(280,220)
\Line(290,60)(290,220)
\Photon(280,140)(320,140){5}{4}
\Photon(360,140)(400,140){5}{4}
\BCirc(340,140){20}
\Text(380,155)[b]{$Z$}
\Text(300,155)[b]{$\gamma $}
\Text(285,55)[t]{$ns_{1/2}$}
\Text(285,225)[b]{$n'p_{1/2}$}
\Text(285,5)[b]{b)}
\Line(395,145)(405,135)
\Line(395,135)(405,145)
\Vertex(285,140){5}
\end{picture}
\end{center}

\                           %Fig.5c vniz
\begin{center}
\begin{picture}(460,230)(0,0)
\Line(60,50)(60,210)
\Line(50,50)(50,210)
\Photon(60,100)(120,100){5}{6}
\Photon(60,150)(100,150){5}{4}
\Photon(140,150)(180,150){5}{4}
\BCirc(120,150){20}
\Line(175,155)(185,145)
\Line(175,145)(185,155)
\Line(115,105)(125,95)
\Line(115,95)(125,105)
\Text(80,165)[b]{$\gamma $}
\Text(160,165)[b]{$\gamma $}
\Text(90,115)[b]{$Z$}
\Text(60,45)[t]{$ns_{1/2}$}
\Text(60,215)[b]{$n'p_{1/2}$}
\Text(55,5)[b]{c)}
\Vertex(55,100){5}
\Vertex(55,150){5}
			   %Fig.5d vniz
\Line(280,50)(280,210)
\Line(270,50)(270,210)
\Photon(280,150)(340,150){5}{6}
\Photon(280,100)(320,100){5}{4}
\Photon(360,100)(400,100){5}{4}
\BCirc(340,100){20}
\Line(395,105)(405,95)
\Line(395,95)(405,105)
\Line(345,155)(335,145)
\Line(345,145)(335,155)
\Text(290,115)[b]{$\gamma $}
\Text(380,115)[b]{$\gamma $}
\Text(310,160)[b]{$Z$}
\Text(280,45)[t]{$ns_{1/2}$}
\Text(280,215)[b]{$n'p_{1/2}$}
\Text(275,5)[b]{d)}
\Vertex(275,100){5}
\Vertex(275,150){5}                 
\end{picture}
\end{center}
\caption{
\label{fig4} 
The Feynman graphs corresponding to the loop corrections in the Uehling 
approximation. The wavy $\gamma$ line with the cross at the end denotes 
the electromagnetic interaction with the nucleus.}
\end{figure}

\newpage
\begin{figure}
\begin{center}
\begin{picture}(460,400)(0,0)
\Line(230,30)(230,300)
\Line(240,30)(240,300)
\PhotonArc(235,165)(50,90,270){5}{22}
\Text(170,165)[l]{$Z$}
\Text(235,315)[b]{$n'p_{1/2}$}
\Text(235,20)[t]{$ns_{1/2}$}
\Vertex(235,215){5}
\Vertex(235,115){5}
\end{picture}
\end{center}
\caption{
\label{5}
The Feynman graph that represents the electron anapole moment correction 
to the basic atomic PNC matrix element in the Furry picture. Notations are 
the same as in Fig. 2.}
\end{figure}
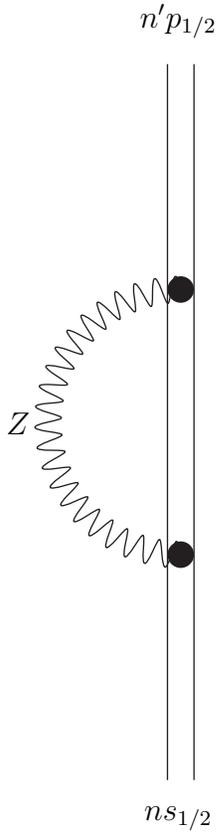

\newpage
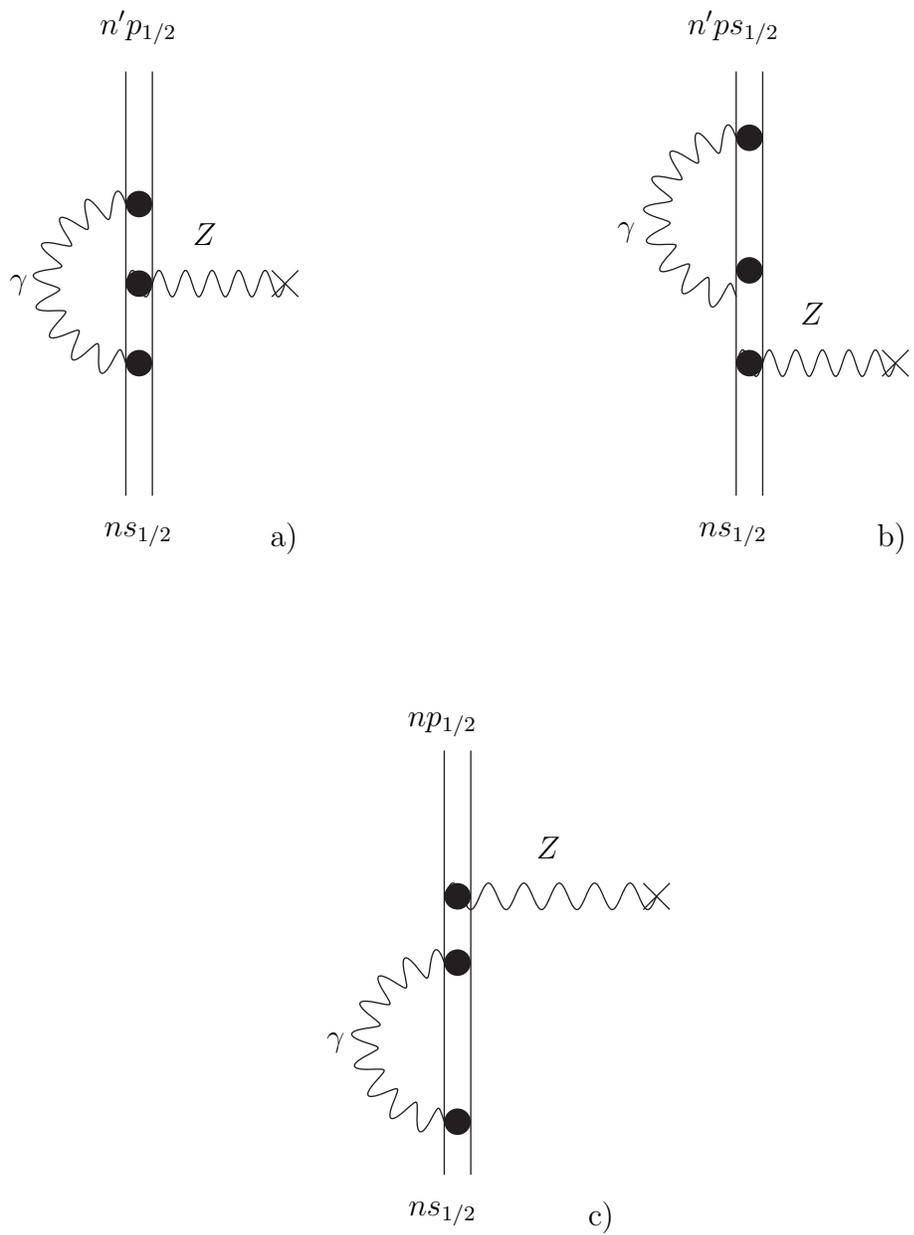
\begin{figure}                          %Fig.9a vniz
\begin{center}
\begin{picture}(460,275)(0,0)
\Line(50,50)(50,210)
\Line(60,50)(60,210)
\Photon(50,130)(110,130){5}{6}
\Line(105,125)(115,135)
\Line(105,135)(115,125)
\PhotonArc(50,130)(30,90,270){5}{9}
\Text(80,145)[b]{$Z $}
\Text(13,130)[r]{$\gamma $}
\Text(55,40)[t]{$ns_{1/2}$}
\Text(55,225)[b]{$n'p_{1/2}$}
\Text(110,40)[t]{a)}
\Vertex(55,130){5}
\Vertex(55,160){5}
\Vertex(55,100){5}
			  %Fig.9b vniz
\Line(280,50)(280,210)
\Line(290,50)(290,210)
\Photon(280,100)(340,100){5}{6}
\Line(335,95)(345,105)
\Line(335,105)(345,95)
\PhotonArc(280,155)(30,90,270){5}{9}
\Text(310,115)[b]{$Z$}
\Text(243,150)[r]{$\gamma $}
\Text(280,40)[t]{$ns_{1/2}$}
\Text(280,225)[b]{$n'ps_{1/2}$}
\Text(340,40)[t]{b)}
\Vertex(285,185){5}
\Vertex(285,135){5}
\Vertex(285,100){5}
\end{picture}
\end{center}
\
			  %Fig.9c vniz
\begin{center}
\begin{picture}(460,233)(0,0)
\Line(170,50)(170,210)
\Line(180,50)(180,210)
\Photon(170,155)(250,155){5}{6}
\Line(245,150)(255,160)
\Line(245,160)(255,150)
\PhotonArc(170,100)(30,90,270){5}{9}
\Text(210,170)[b]{$Z $}
\Text(133,100)[r]{$\gamma $}
\Text(170,40)[t]{$ns_{1/2}$}
\Text(170,220)[b]{$np_{1/2}$}
\Text(230,40)[t]{c)}
\Vertex(175,130){5}
\Vertex(175,70){5}
\Vertex(175,155){5}
\end{picture}
\end{center}
\caption{ \label{6}
Feynman graphs that represent the corrections caused by the 
electromagnetic renormalization of the $Z$ coupling. }
\end{figure}

\newpage
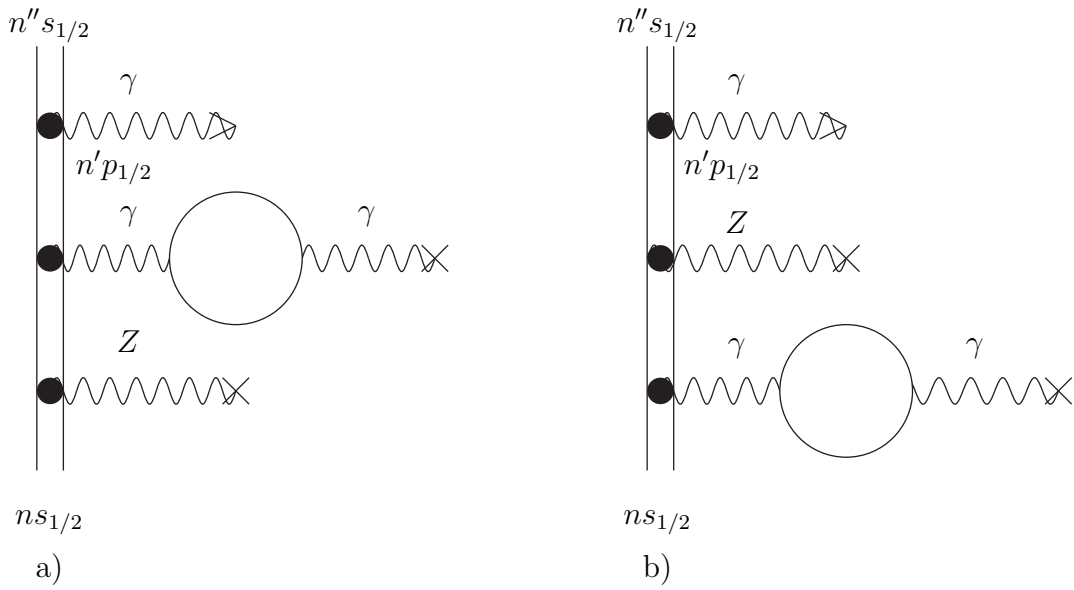
\begin{figure}                               %Fig.10a vniz
\begin{center}
\begin{picture}(460,240)(0,0)
\Line(50,50)(50,210)
\Line(60,50)(60,210)
\Photon(55,80)(125,80){5}{7}
\Photon(55,180)(125,180){5}{7}
\Photon(55,130)(100,130){5}{5}
\Photon(150,130)(200,130){5}{5}
\Line(120,85)(130,75)
\Line(120,75)(130,85)
\Line(115,185)(125,180)
\Line(115,175)(125,180)
\Line(195,135)(205,125)
\Line(195,125)(205,135)
\Vertex(55,80){5}
\Vertex(55,180){5}
\Vertex(55,130){5}
\Text(55,35)[t]{$ns_{1/2}$}
\Text(55,225)[t]{$n''s_{1/2}$}
\Text(65,165)[l]{$n'p_{1/2}$}
\Text(85,95)[b]{$Z$}
\Text(85,145)[b]{$\gamma $}
\Text(175,145)[b]{$\gamma $}
\Text(85,195)[b]{$\gamma $}
\Text(55,10)[b]{a)}
\BCirc(125,130){25}
			       %Fig.10b vniz
\Line(280,50)(280,210)
\Line(290,50)(290,210)
\Photon(285,80)(335,80){5}{5}
\Photon(285,180)(355,180){5}{7}
\Photon(375,80)(435,80){5}{5}
\Photon(280,130)(355,130){5}{7}
\Line(350,125)(360,135)
\Line(350,135)(360,125)
\Line(345,185)(355,180)
\Line(345,175)(355,180)
\Line(430,85)(440,75)
\Line(430,75)(440,85)
\Vertex(285,80){5}
\Vertex(285,180){5}
\Vertex(285,130){5}
\Text(285,35)[t]{$ns_{1/2}$}
\Text(285,225)[t]{$n''s_{1/2}$}
\Text(295,165)[l]{$n'p_{1/2}$}
\Text(315,95)[b]{$\gamma $}
\Text(315,140)[b]{$Z $}
\Text(405,95)[b]{$\gamma $}
\Text(315,195)[b]{$\gamma $}
\Text(285,10)[b]{b)}
\BCirc(355,80){25}
\end{picture}
\end{center}
\caption{\label{fig7} 
Feynman graphs for the amplitude of the process of the photon emission including PNC. } 
\end{figure}

\newpage
\begin{figure}
\begin{center}
\begin{picture}(460,240)(0,0)
\Line(150,50)(150,210)
\Line(160,50)(160,210)
\Photon(155,130)(225,130){5}{7}
\Photon(155,80)(225,80){5}{7}
\Photon(155,180)(200,180){5}{5}
\Photon(250,180)(300,180){5}{5}
\Line(220,125)(230,130)
\Line(220,135)(230,130)
\Line(220,85)(230,75)
\Line(220,75)(230,85)
\Line(295,185)(305,175)
\Line(295,175)(305,185)
\Vertex(155,80){5}
\Vertex(155,180){5}
\Vertex(155,130){5}
\Text(155,35)[t]{$ns_{1/2}$}
\Text(155,225)[t]{$n''s_{1/2}$}
\Text(165,110)[l]{$n'p_{1/2}$}
\Text(185,95)[b]{$Z$}
\Text(185,195)[b]{$\gamma $}
\Text(275,195)[b]{$\gamma $}
\Text(185,145)[b]{$\gamma $}
\BCirc(225,180){25}
\end{picture}
\end{center}
%\begin{center}
%Fig. 8
%\end{center}
\caption{
\label{fig8}  
Feynman graph  that describes the radiative corrections to the emission 
matrix element.}
\end{figure}
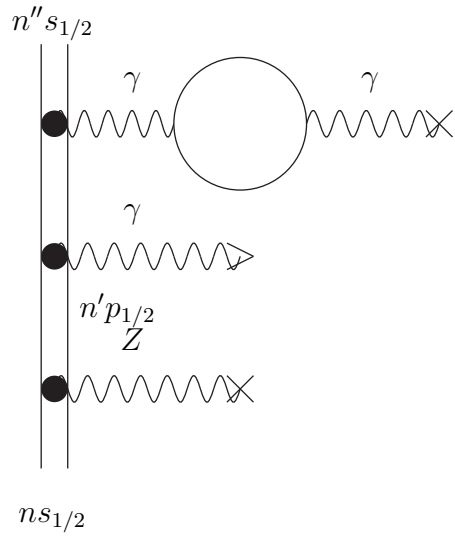

\newpage      
\begin{thebibliography}{99}
\bibitem{l1} Review of Particle Properties, Particle Physics Booklet, AIP, July 1996
\bibitem{l2} M.~S.~Peskin and T.~Takeuchi, Phys.~Rev. D46, 381 (1992)
\bibitem{l3} A.~Sirlin, Phys.~Rev. D22, 971 (1980)
\bibitem{l4} W.~J.~Marchiano and J.~L.~Rosner, Phys.~Rev.~Lett. 65, 2963 (1990) \bibitem{l5} K.~T.~Mahanthappa and P.~K.~Mohapatra, Phys.~Rev D43, 3093 (1991) 
\bibitem{l6} B.~W.~Lynn and P.~G.~H.~Sandars, J.~ Phys. B27, 1469 (1994)
\bibitem{l7} M.~A.~Bouchiat and C.~C.~Bouchiat, Phys.~Lett. 48b, 111 (1974)
\bibitem{l8} J.~Schweppe, A. Belkacem, L. Blumenfield, N. Claytor, 
B. Feinberg, H. Gould, V. E. Kloster, L. Levy, S. Misawa, 
J. R. Mowat, M. H. Prior, Phys.~Rev.~Lett. 66, 1434 (1989)

\bibitem{l9} Th. St\"ohlker, P. H. Mokler, K. Beckert, F. Bosch, 
H. Eickhoff, B. Franzke, M. Jung, T. Kandler, O. Kleppner, 
C. Kuzhuharov, R. Moshammer, F. Nolden, H. Reich, P. Rymuza,
P. Sp\"adtker, M. Steck, Phys.~Rev.~Lett. 71, 2184 (1993)

\bibitem{l10} H. F. Beyer,  
G. Menzel, D. Liesen, A. Gallus,
F. Bosch, R. D. Deslattes, P. Indelicato, Th. St\"ohlker, 
O. Kleppner, R. Moshammer, F. Nolden, H. Eickhoff, B. Franzke,
M. Steck,  
Z.~Phys. D35, 169 (1995)

\bibitem{l11} H. F. Beyer, D. Liesen, F. Bosch, K. D. Finlayson, 
M. Jung, O. Kleppner, R. Moshammer, K. Beckert, H. Eickhoff, B. Franzke,
F. Nolden, P. Sp\"adtker, M. Steck, G. Menzel, R. D. Deslattes, 
Phys.~Lett. A184, 435 (1994) 
\bibitem{l12} Th. Beier and G. Soff, Z.~Phys. D8, 129 (1988)
\bibitem{l13} S. M. Schneider, W. Greiner and G. Soff, 
J. Phys. B26, L529 (1993)  
\bibitem{l14} I. Lindgren, H. Persson, S. Salomonson, V. Karasiev, L.~Labzowsky, A.~Mitruschenkov and M.~Tokman,
J.~Phys. B26, L503 (1993)
\bibitem{l15} A. Mitrushenkov, L. Labzowsky, I. Lindgren, 
H.~Persson and S.~Salomonson,
Phys.~Lett. A200, 51 
(1995)
\bibitem{l16} H. Persson, I. Lindgren, L. Labzowsky, G.~Plunien, T.~Beier 
and G.~Soff,
Phys.~Rev. A54, 2805 (1996) 
\bibitem{l17}  S. Mallampalli and J. Sapirstein, Phys.~Rev. A54, 2714 (1996)
\bibitem{l18} P. G. H. Sandars, J. Phys. B23, L655 (1990) 
\bibitem{l19} L. Labzowsky, G. Klimchitskaya and Yu. Dmitriev, 
{\em Relativistic Effects in Atomic System}, 
(IOP, Bristol and Philadelphia, 1993)
\bibitem{l20} N. N. Bogolyubov and D. V. Shirkov, {\em Quantenfelder}, 
(Physik-Verlag, Weinheim, 1985)
\bibitem{l21} M. Gell-Mann and F. Low, Phys. Rev. 84, 350 (1951)
\bibitem{l22} J. Sucher, Phys. Rev. 107, 1448 (1957)
\bibitem{l23} L. Labzowsky, V. Karasiev, I. Lindgren, H. Persson and
S. Salomonson, Phys. Scripta T46, 150 (1993)
\bibitem{l24} A. Sch\"afer, G. Soff, P. Indelicato, B. M\"uller and W. Greiner,
Phys. Rev. A40, 7362 (1989)
\bibitem{l25} V. V. Karasiev, L. N. Labzowsky and A. V. Nefiodov, 
Phys. Lett. A172, 62 (1995)
\bibitem{l26} E. A. Uehling, Phys. Rev. 48, 55 (1935)
\bibitem{l27} A. Akhieser and V. Berestezkii, {\em Quantum Electrodynamics}, 
(Wiley, New York, 1965)
\bibitem{l28} F. Salvat and R. Mayol, Comput. Phys. Commun. 62, 65 (1991)
\bibitem{l29} R. W. Dunford, Phys. Rev. A54, 3820 (1996)   
\end{thebibliography}
\end{document}